# Study of participant-spectator matter formed in a heavy-ion collision

Sukhjit Kaur[1], Aman D. Sood[2] and Rajeev K. Puri[1*]

[1]*Physics Department, Panjab University Chandigarh -160014, INDIA*

[2]*SUBATECH,*

*Laboratoire de Physique Subatomique et des Technologies Associées*

*Université de Nantes - IN2P3/CNRS - EMN*

*4 rue Alfred Kastler, F-44072 Nantes, FRANCE*

*\*email: rkpuri@pu.ac.in*

## Introduction

Heavy-ion collisions at intermediate energies offer an opportunity of producing nuclear systems under the extreme conditions of temperature and density. At high excitation energies, the colliding nuclei may break-up into several small and intermediate size fragments followed by the emission of large number of nucleons. The fragmentation of colliding nuclei into several pieces of different sizes is a complex phenomenon. This may be due to the interplay of the correlations and fluctuations emerging in a collision. Several studies, in literature, have been made to check the fragmentation pattern. It is found that fragmentation pattern depends on the size of the colliding nuclei, incident energy as well as impact parameter [1,2]. Sood and Puri [3] checked the system size dependence of average and maximum central density, temperature, collision dynamics, and participant and spectator matter, as well as the time zone for hot and dense nuclear matter at the energy of vanishing flow and observed a power law dependence of various quantities. So far, no study exists, in the literature, at the energy of maximal production of intermediate mass fragments (IMFs). Here, we plan to the check system size dependence of all these quantities at energies at which the maximal production of IMFs occurs.

This study is made within the framework of quantum molecular dynamics model, which is described in detail in Refs. [4-6].

## Results and Discussion

We studied the central reactions of $^{20}Ne+^{20}Ne$, $^{40}Ar+^{45}Sc$, $^{58}Ni+^{58}Ni$, $^{86}Kr+^{93}Nb$, $^{129}Xe+^{118}Sn$, $^{86}Kr+^{197}Au$ and $^{197}Au+^{197}Au$ using hard equation of state along with Cugnon cross section. The incident energies used in the above reactions are the energies at which the maximal production of IMFs occurs. These energies read as 23.7, 45.7, 69.3, 77.6, 96.2, 124.4, and 104.5 AMeV, respectively, for above mentioned systems. All reactions are followed till 200 fm/c.

In the present study, we have checked the mass dependence of various quantities such as participant and spectator matter, average and maximum central density, collision dynamics as well as the time zone for hot and dense nuclear matter.

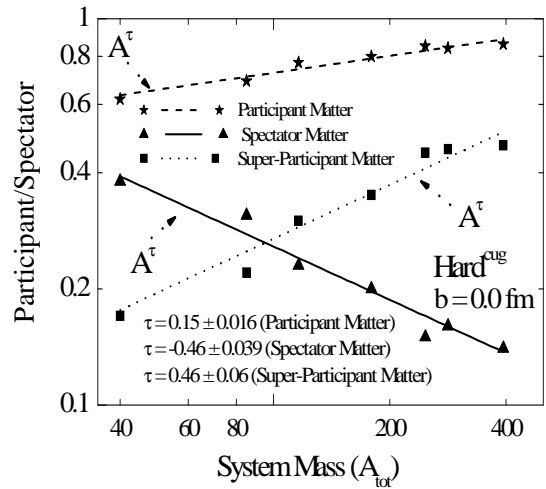

**Fig.** 1 : The participant, spectator, and super-participant matter (defined in terms of nucleon-nucleon collisions) as a function of composite mass of the system. The dashed, dotted, and solid lines represent the $\chi^2$ fits with power law $cA_{tot}^{\tau}$.

In fig. 1, we display participant, spectator, and super-participant matter (obtained at 200 fm/c and defined in terms of nucleon- nucleon c-ollisions) as a function of the total mass of the system. All nucleons that have experienced at least one collision are termed as participant matter. The remaining matter is counted as spectator matter. The nucleons having experienced more than one collision are termed as super-participant matter. As evident from fig.1, participant matter shows nearly mass independent behavior ($\tau = 0.15 \pm 0.016$). The spectator and super-participant matter, however, exhibit a power law dependence with $\tau = -0.46 \pm 0.039$ for spectator matter and $\tau = 0.46 \pm 0.06$ for super-participant matter.

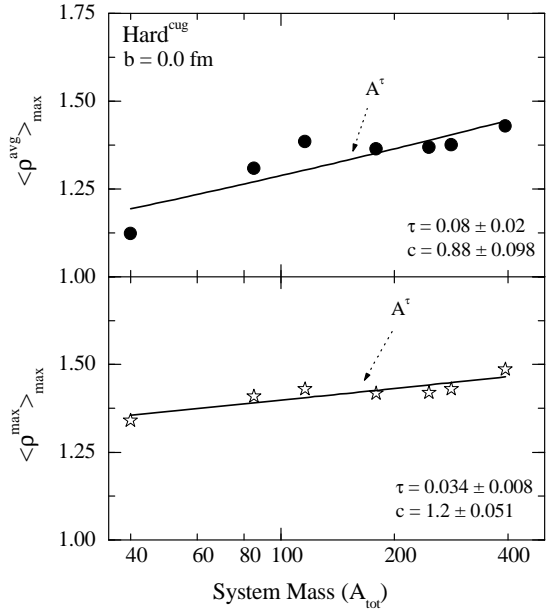

**Fig. 2**: The maximal value of the average density $\langle\rho^{avg}\rangle_{max}$ (upper part) and maximum density $\langle\rho^{max}\rangle_{max}$ (lower part) as a function of the composite mass of the system. The solid lines are the fits to the calculated results using $cA_{tot}^{\tau}$ obtained with $\chi^2$ minimization.

In fig. 2, we display the system size dependence of the maximal value of $\langle\rho^{avg}\rangle$ and $\langle\rho^{max}\rangle$. Most of the reactions considered here are nearly symmetric in nature (except $^{86}Kr+^{197}Au$). The maximal values of $\langle\rho^{avg}\rangle$ and $\langle\rho^{max}\rangle$ follow a power law proportional to $A_{tot}^{\tau}$ with $\tau$ being $0.08 \pm 0.02$ for the average density, $\langle\rho^{avg}\rangle$ and $0.034 \pm 0.008$ for maximum density, $\langle\rho^{max}\rangle$. Alternatively, we can say that a slight increase in the density occurs with increase in the size of the system. If all these reactions simulated at a fixed value of energy, an entirely different trend is noted.

It is worth mentioning that the collision dynamics as well as the time zone for hot and dense nuclear matter also show power law dependence.

## Acknowledgments

This work is supported by Indo-French Project.